# GLOBAL TRANSFERS: M-PESA, INTELLECTUAL PROPERTY RIGHTS AND DIGITAL INNOVATION

Christopher Foster, University of Manchester, christopher.foster-2@manchester.ac.uk

**Abstract:** In July 2020, in the midst of the COVID crisis, the Kenyan mobile operator Safaricom announced that the intellectual property rights (IPR) for mobile money service M-Pesa were "moving back into African control". This paper tracks how the IPR originally came to be held outside Kenya, and the implications for understanding M-Pesa as an inclusive innovation. Through reflection of this analysis of IPR and innovation, the paper contributes to discussions on structural aspects of digital innovation in the global south. By focussing on IPR, it unpacks some of the processes by which global intellectual property regimes and cross-border IPR practices shape uneven outcomes and power.

## 1. INTRODUCTION

There has been an emergence of global digital innovations that follow common patterns - operating at a global scale, owned by large firms and shaped by demands of financialization, market dominance, intellectual property control and data ownership (Srnicek, 2016). A crucial question is how to integrate these more critical perspectives into research on digital inclusion and development. Literature on inclusive innovation has made pragmatic investigations of how technologies and processes might be inclusive of marginalised groups (Heeks et al., 2014). But the digital products and services being discussed are frequently the very same ones that are being critiqued as part of the emerging era of 'digital capitalism'.

This paper focuses on one aspect of this issue by exploring intellectual property rights (IPR) in digital innovation. Who gains rights when intellectual property is generated is a key structuring aspect in shaping modern economies, and plays an important role in determining who gains from innovation. Exploring the role of IPR in the contexts of digital innovations can therefore provide important perspectives on digital development.

This paper explores the IPR through the case of Kenyan mobile money service M-Pesa. M-Pesa has been fraught with tensions around IPR during its entire history. These tensions can be argued to have had a significant influence on M-Pesa trajectories and its impacts. The paper highlights how M-Pesa is part of a global network of actors with different agendas and with complex relationships of control. In the case of M-Pesa, not only has control of IPR led to significant bottlenecks in innovation in Kenya, but also significant financial transfers to the UK which call into question the more optimistic claims of inclusive innovation.

## 2. INNOVATION AND IPR AT A MACRO LEVEL

### 2.1. Innovation in the global south

Subsets of digital innovation linked to development goals initially tended to follow paradigms linked to ICT for development (ICT4D) (Unwin, 2009). Small-scale digital innovations and infrastructure were rolled out in specific contexts, supported by diverse actors including governments, NGOs, corporate CSR and telecoms firms (Murphy & Carmody, 2015). With the growth of access, however, it has been useful to think about digital innovation as having moved away from ICT4D towards "digital development", aligned to the expanding digital economy across the globe (Heeks, 2020). As such increasingly one needs to consider the relationship between global firms involved in the digital economy and claims about impacts in the global south.

An important framework for examining such relationships is that of inclusive innovation which examines "the means by which new goods and services are developed for and by marginal groups" (Foster & Heeks, 2015, p. 2). The discussion of development impacts has often come through a focus on 'inclusivity' examining local adoption, use and small-scale impacts (Heeks et al., 2013). More systematic thinking has, however, begun to move beyond such fleeting development gains to think more systematically about how innovations reproduce or change processes within and across societies in areas such as innovation and gender, neoliberal growth models and the ownership of knowledge (Jiménez & Zheng, 2017; Pansera & Owen, 2016; Papaioannou, 2016; Smith et al., 2016). This paper aligns to such work through an analysis of intellectual property, an important structuring factor in the economy. In the next sections, some of the key tenets of IPR institutions are discussed to understand their links to innovation in the global south.

## 2.2. IPR and inclusive innovation

"Intellectual property rights" include a range of different rights that protect individuals and organisations who create "intellectual property". This includes copyright on creative works and computer programs, trademarks on brands and marks, and rights on innovative ideas including patents, industrial designs and trade secrets.

Underlying global regimes of intellectual property are powerful global institutions, particularly the World Intellectual Property Organisation (WIPO) and the Trade-Related Aspects of Intellectual Property Rights (TRIPS) that define strong intellectual property regimes across the globe (Peukert, 2017). Three decades after the introduction of TRIPS, the evidence on strong IPR regimes suggests overall negative consequences towards the global south (Chang, 2001; Peukert, 2017; Stiglitz, 2014). Those examining global trade have critiqued the impacts of IPR regimes which narrow the ability for global south states to build domestic innovation and tend to reinforce the dominance of IPR within the global north (Chang, 2001; Lall, 2003).

Over time a range of more detailed critiques of strong IPR regimes have been discussed that are relevant to inclusive innovation. Important questions have emerged about how societal orientated research can fit into IPR regimes that primarily concentrate on firm ownership (Ghauri & Rao, 2009). Organisations that are involved in innovation (non-profit, public, universities, social enterprises) may have agendas that go beyond profit, but these goals can often be dissipated within the present institutions. Furthermore, as IPR has become more global, a key issue relates to differentiating between firm created intellectual property and local and indigenous knowledge, creativity and innovation. Case studies have highlighted these concerns in the global south, when indigenous knowledge 'discovered' by corporate actors are problematically incorporated into intellectual property (Finger & Schuler, 2004; Srang-iam, 2013).

## 2.3. IPR and the digital economy

Intangibles, investments made in creating intellectual assets, are increasingly important in an economy driven by information (Haskel & Westlake, 2017). This is particularly the case for global digital firms, where the generation of intellectual capital (ideas, brands, software, data, relationships) are at the core of their global business models. As firms move beyond material assets, IPR is a crucial way to protect these investments. WIPO data has highlighted growth in IPR in recent years: with global patents growing from 1.6m filings in 2004 to 3.3m in 2018; and trademarks applications from 4.6m in 2004 to 14.3m in 2018 (WIPO, 2020). For technology-intensive firms, building 'portfolios' of intellectual property claims is now an important part of firm strategy in competitive marketplaces (Rice, 2015). These expansionist trajectories of IPR have also emerged out of the relaxation of requirements of novelty in patents. Increasingly, patents based on software and business models have been permitted in many nations, when previously they were seen as abstract ideas and therefore less patentable (Bessen & Hunt, 2007; Hall & Harhoff, 2012).

IPR assigns firm exclusivity to an ever-growing range of ideas, brands and processes this has led to the argument that it takes ideas and innovation away from the 'commons' with important economic impacts (Stiglitz, 2014). From a development perspective, these trends are highly problematic considering the often-global scope on IPR and the skew in IPR ownership towards the global north (ibid). Some accounts of IPR expansionism suggest that we are seeing a small number of firms building out global intellectual monopolies, where "intellectual monopoly capitalism" is defining a new domain of global economic dominance beyond markets (Pagano, 2014). A key example of intellectual monopoly capitalism has been global digital firms that have been at the forefront of IPR expansionism and rapid globalisation (Durand & Milberg, 2020; Rikap, In press).

In summary, this section outlined some of the important aspects of innovation and intellectual property. It introduced the link between inclusive innovation and the global digital economy. It outlined the emergence of strong IPR regimes that are skewed towards the global north. It has highlighted the growing importance of intangible assets in the digital economy and the implications of IPR expansionism. Such arguments are well documented at this macro-level, but there is very little analysis of how such processes unfold on the ground. In particular, little is known about how these processes shape innovative digital products and services in the global south where development aims (or claims) are included.

## 3.    APPROACH

The goal of the remainder of the paper is to unpack these issues through an analysis of M-Pesa. The paper undertakes a qualitative analysis examining the history of IP and IPR. A qualitative approach is crucial as it allows us to unpack the specific activities that have had significant effects on IPR.

The analysis builds upon outcomes of qualitative research undertaken with a range M-Pesa actors in Kenya during the periods 2010-2012 and 2015-2016. This provides the foundation for understanding the historic evolution of the service. Alongside this, the author has continued to be involved in conversations and workshops with various policy and practice actors, including M-Pesa agents and ICT consultants[1]. In addition, this paper analyses a range of technical documentation relevant to the IPR including patents, project evaluations, and company documentation which are key to illuminating the often-technocratic aspects of intellectual property.

The case is set up as follows. In section 4.1 the paper provides an overview of the case, introducing the M-Pesa innovation and actors, and summarising the overall impacts of the IPR. Section 4.2 then makes a detailed historical account of how intellectual property has been generated and the different aspects of intellectual property rights. For space, only the two most important aspects of IPR are discussed - patents and software.

---

[1] Given that this paper will argue that intellectual property is multi-faceted and complex, it is appropriate for the author to acknowledge that this paper has been inspired by a vibrant discussion amongst interviewees and online in Kenya over the last decade. This has often centered on the debate "is M-Pesa a Kenyan innovation?". Interview respondents and those within Kenyan media and blogosphere have made interesting interventions to this discussion which are relevant to this analysis of intellectual property. This paper therefore acknowledges the work done by Alex Kuria, Tefo Mohapi, Brian Gachichio, Cyprian Nyakundi, Kenya West and Nairobi Law Monthly

# 4.  M-PESA AND IPR

## 4.1.  Introducing M-Pesa

### 4.1.1.  Key actors

Accounts of the M-Pesa service has been outlined in great detail elsewhere (e.g. Foster & Heeks 2013a, Hughes & Lonie 2007, Ngugi et al. 2010, Omwansa 2009) therefore only a brief account is reprised in this paper relevant to the history of IPR. M-Pesa is a well-known service in Kenya that initially allowed money to be exchanged between mobile phone users through mobile messaging. It provides a convenient mechanism for person-to-person transfer of money, with significant use by low-income groups. Over time, M-Pesa has expanded to serve as a platform for an increasingly diverse range of services as outlined in the most recent statistics in Table 1.

|  | Total for Quarter | M-Pesa | M-Pesa market share |
|---|---|---|---|
| Active subscriptions | 29,185,577 | 28,842,584 | 98.8% |
| Agents | 202,102 | 173,259 | 85.7% |
| Transfers |  |  |  |
| *Business-to-business* | $8,360,196,019 | $8,360,196,019 | 100.0% |
| *Person-to-person* | $6,417,217,788 | $6,413,246,371 | 99.9% |
| *Business-to-consumer* | $3,477,971,699 | $3,475,750,580 | 99.9% |
| *Consumer-to-business* | $2,955,776,655 | $2,952,232,096 | 99.9% |
| *Consumer-to-gov* | $221,914,717 | $108,827,318 | 49.0% |
| Deposits | $5,789,785,352 | $5,774,508,633 | 99.7% |

**Table 1: Mobile money statistics in Kenya, Q3 2019-2020 (Jan – March 2020)**
**Source: (CAK, 2020) based on operator returns.**

M-Pesa is run by Kenyan mobile operator Safaricom. The beginnings of the M-Pesa service emerged in a project supported by the UK's Department of International Development (DFID) with DFID provided matched funding alongside UK-based Vodafone Group, the parent company of Safaricom. The basis of the M-Pesa service was an outcome of this project. The story of the rapid scaling of M-Pesa in Kenya is well-known, with M-Pesa revenue becoming an increasingly important aspect of Safaricom's financial health as outlined in Figure 1.

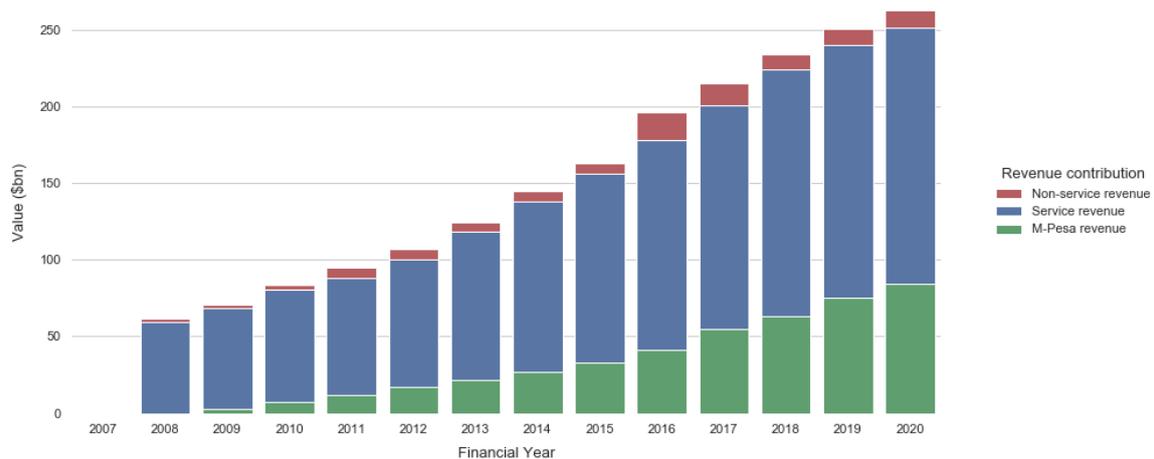

**Figure 1: M-Pesa proportion of revenue**
**Source: Author's calculations based on Safaricom financial reports.**

Several Kenyan actors play an important role in the service. Agents are independent firms who provide the local "face of M-Pesa" through registration and cash-conversion service on the ground. Over time, the number of third parties involved in M-Pesa has also greatly expanded as it has become a platform, integrating a range of banks, business, micro-finance and government actors.

Table 2 highlights corporate actors associated with Safaricom. In Kenya, the government retains a large stake in Safaricom, as well as significant shareholdings, said to be owned by elite groups (Tyce 2020). Thus, while the UK's Vodafone Group has controlling ownership of Safaricom, the firm has a range of shareholders, sometimes with diverging agendas related to their domestic or international focus.

One important driver of direction for Vodafone Group is that it controls or has stakes in a range of mobile operators across the globe. With M-Pesa's success, Vodafone Group has leveraged the service by introducing it across a range of its subsidiaries including Egypt, Ghana, DRC, Tanzania, Mozambique, Lesotho and India (Vodafone Group, 2019).

| Period | Ownership of Safaricom |
|---|---|
| Pre-IPO (June 2008) | 60% Government of Kenya<br>40% Vodafone Kenya Limited<br>● *87.5% Vodafone Group, 12.5% Mobiletea* |
| Post-IPO | 35% Government of Kenya<br>40% Vodafone Kenya Limited<br>● *100% Vodafone group*<br>25% Market Shareholders via IPO |
| Post-Vodafone restructure (2017) | 35% Government of Kenya<br>40% Vodafone Kenya Limited<br>● *87.5% Vodacom, S Africa, 12.5% Vodafone Group*<br>25% Market Shareholders |

**Table 2: Major ownership changes in Safaricom**
**Source: Safaricom annual reports and company presentations**

An important driver of recent changes of M-Pesa IPR assets is the 2017 move of Safaricom to come under the control of Vodacom, South Africa (itself a subsidiary of Vodafone Group UK) (Vodafone Group, 2017). Official explanations for this corporate reshuffle revolve around regional simplification of structures (Vodafone Group, 2017), but it has also been reported in the business press that this was done with a possible sell-off of Vodafone's African assets in mind (Fildes, 2017).

### 4.1.2. Impacts of IPR

Key intellectual property rights related to M-Pesa include its trademarks, patents related to innovative aspects of the service and the copyright embedded in the software for the service. From its launch in 2007 up until 2020, M-Pesa IPR was controlled by the Vodafone Group. To use the M-Pesa service Safaricom, therefore, paid an annual fee to the Vodafone Group (as shown in Table 2). This was positioned as a "service fee" that allowed Safaricom to use the M-Pesa service (and permitted the use of other IPR assets).

The direct impact of the service fee has been significant transfers from Safaricom to Vodafone Group. Figure 2 & Table 3 show an estimate of the value of these service fees, which reached a peak of around €26m in 2015. As an estimate for the period 2010-2019, this resulted in a direct transfer of around €170m. One might find the extent of such transfers surprising given the origins of M-Pesa as a jointly funded development project and the development claims that have been made over the years about the service.

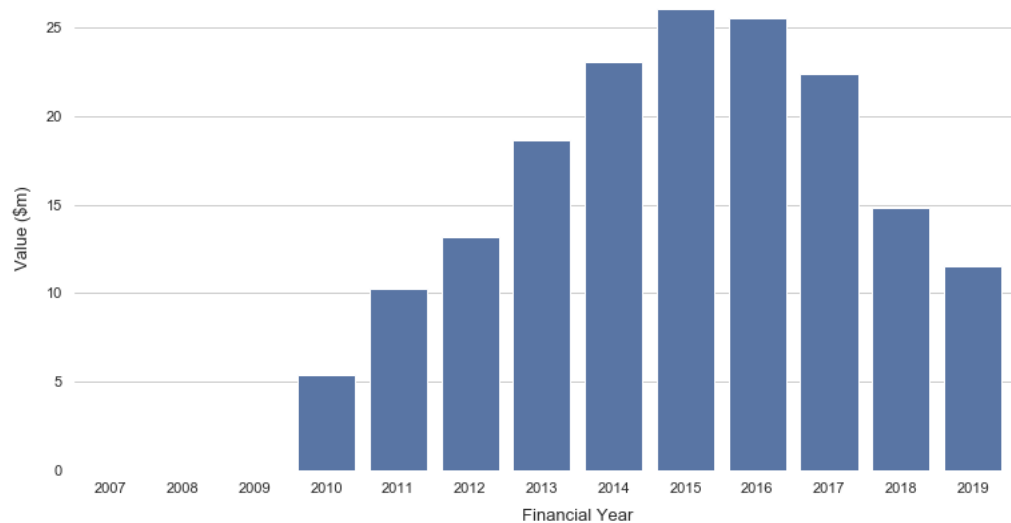

**Figure 2: Transfers from M-Pesa. See Table 3 for details**
**Source: Author's calculations based upon company reports**.

| Financial Year | 10 | 11 | 12 | 13 | 14 | 15 | 16 | 17 | 18 | 19 |
|---|---|---|---|---|---|---|---|---|---|---|
| M-Pesa revenue (€m) | 72.0 | 99.6 | 152.5 | 198.7 | 222.4 | 327.7 | 359.4 | 501.4 | 505.0 | 663.2 |
| M-Pesa service fee (%) | 10 | 10 | 10 | 10 | 10 | 10 | 10/5* | 5 | 2 | 2 |
| Estimated service payment (€m) | 5.3 | 10.2 | 13.2 | 18.6 | 23.1 | 26.1 | 25.6 | 22.4 | 14.9 | 11.6 |
| * Note: Fee moves from 10% to 5% on 1st Aug 2015 | | | | | | | | | | |

**Table 3: M-Pesa revenues and transfers**
**Source: Author's calculations based upon company reports. See appendix for discussion.**

It is also important to highlight the indirect impacts that IPR control has had on innovation trajectories. In interviews with those involved in M-Pesa in Kenya, there has been an ongoing frustration for many years about the relatively slow upgrading of new services and features of M-Pesa (Foster, 2014), with Safaricom tending to prefer simple integration of new actors over radical changes. One of the key reasons for this is that with Vodafone Group controlling the IPR, and with the use of M-Pesa across the globe it has become more difficult for Safaricom to influence the directions around the service and push technical upgrading.

This may change with the new joint venture on M-Pesa announced in July 2020 between Safaricom and Vodacom South Africa. As mentioned in the abstract, the joint venture was reported in the press as bring back "full control of the M-Pesa brand, product development and support services" to Kenya (Phillip, 2020), although what this means in terms of the ownership of IPR has not yet been spelt out. The two firms each paid 1,073million KSh (£7.02m) to form the Joint Venture. Associated with this has been strong articulations that it will lead to a renewal in innovation of the M-Pesa service in Africa (Vodacom, 2019).

### 4.2. Unpacking M-Pesa's IPR

#### 4.2.1. The origins of IPR

As outlined in the previous section, the origin of M-Pesa was a donor project for the UK development agency DFID. It is worth examining the treatment of IPR issues in this project as an important aspect of the history of the IPR.

The project occurred under the DFID Financial Deepening Challenge Fund (FDCF), which supports commercial firms to become more involved in development. There have been significant evaluations made around such donor "challenge funds". Evaluations suggest they are often "launched with more enthusiasm than forethought…there is considerable confusion about what international development agencies should be seeking to achieve by means of them, and how" (Davies & Elgar, 2014). The typical focus was that they would support private firms by making certain markets more financially viable in the global south (Brain et al., 2014). With comparatively limited amounts being provided, and typically structured as one-off funding, there has rarely been an expectation of innovation (Pompa, 2013).

With innovation being a peripheral goal, approaches to dealing with IPR have been poorly thought out. When a specific project has become innovative (such as M-Pesa) there appears to be a lack of clear processes to deal with issues. Internal reports and reviews of the FDCF have alluded to such issues. In the mid-term review, IPR is brought up as a significant problem.

*"Intellectual property rights were also not adequately considered in the design process. As many proposals involve customised ICT development, this poses a serious problem" (Ebony Consulting, 2003, p. 9)*

A principle in donor-supported projects in the UK is that IPR should be shared, where firms would "give DFID a license to use the project material; this would include the right to publish, copy and distribute all material include written material and software produced under a FDCF grant" (Ebony Consulting, 2003, p. 28). This, however, proved problematic where some commercial firms suggested that "signing an agreement that potentially gave DFID rights to their IPR was a problem" (Irwin & Porteous, 2005, p. 9). In the case of the FDCF, some workarounds were made,

*"… require IPR co-ownership only on software developments made under grant funding. That means that if the initial software was acquired before the grant, the institution will retain exclusive IPR" (Ebony Consulting, 2003, p. 28).*

Overall, in reflecting on such experiences, the full-term review of the FDCF ultimately suggested that the best way forward on IPR was to

*"avoid funding the development of new technology...it avoids getting into debates about who ultimately owns the IPR" (Irwin & Porteous, 2005, p. 44).*

The recommendations outlined here are embedded in long unread institutional reports and might seem minor. But the direction taken in the early 2000s underlies the IPR tensions that face in M-Pesa today. The lack of guidance and workarounds in the FCDF contributed to M-Pesa IPR becoming solely owned and controlled by the parent company Vodafone Group in the UK.

### 4.2.2. Contributors to intellectual property

In contrast to this singular ownership and control of IPR, a range of actors have played an important role in shaping the M-Pesa service. This discussion is not intended to suggest that Vodafone or Safaricom have not played a key role, but it highlights a diverse set of contributors.

The key contributors are summarised in Table 4. Examples of those contributing include those involved in precursor innovation such as those transferring money more informally within East Africa using mobiles which were well-known prior to pilot studies. Trial users in the donor pilot project also played a crucial role in shaping the M-Pesa service to what it is today. As the service was rolled out, agents and other associates have driven forward the service through adaptions as it has rapidly scaled.

| Activity | Details and actors | Source |
| --- | --- | --- |
| Precursor innovations | M-money services such as SmartMoney (2000) G-Cash (2004) (Philipines) and CelPay (2002, Zambia) pre-date M-Pesa with very similar models.<br><br>Informal money transfer using mobile happening during the early rollout of mobile money in East Africa approximating M-Pesa | (Chipchase & Tulusan, 2006; Porteous, 2006) |
| DFID trial | The DFID trial was originally positioned as a broader micro-finance project with the core transfers of M-Pesa are just one component.<br><br>Accounts of the pilot outline how pilot participants (users and agents) began to undertake mobile transfers that were outside the principal micro-finance activity. This was identified by the pilot coordinators and drove the idea that the main focus should solely be on the mobile transfers | (Foster & Heeks, 2013b; Hughes & Lonie, 2007) |
| Policy actors in commercialisation | A key process post-trial before commercial rollout was ensuring the service fit within regulation. This involved inputs from several players including the UK Financial Services Authority (FSA) and the Central Bank of Kenya (CBK) | (AFI, 2010; Morawczynski, 2010) |
| Agents and scaling | Once the service was launched, adaptations of the service were a key to service growth, involving a diversity of actors including agents and other associates. These activities were crucial in ensuring that the service was viable and manageable as it rapidly scaled. | (Foster & Heeks, 2013b) |

**Table 4: Diversity of actors contributing to M-Pesa.
Source: Extended from (Foster, 2013)**

These contributors to M-Pesa align with the idea that innovation should be seen as a process. Such a position closely connects with theories of "innovation systems" that see the processes of innovation as increasingly distributed amongst an interacting set of actors (see extended discussion in Foster & Heeks, 2013a).

### 4.2.3. Patents

An important part of M-Pesa IPR is a set of patents that have emerged over the last decade (listed in Table 5). The patent profile associated with M-Pesa has primarily been defensive so far[2]. We examine three aspects of the 10 patent filings found, in line with the literature discussion on IPR[3]: the origins of the ideas in these patents, the basis of novelty claims, and the type/scope of the patent.

---

[2] These patents appear to have been little used in litigation at present, besides some cases in Kenya where filing dates have enabled Safaricom to claim prior art over competing claims of invention on specific aspects of mobile money (e.g. Mohapi, 2017)

| Title | Priority Date | Area |
|---|---|---|
| 1) Mobile Account Management (Murdoch et al., 2007) | 2005 | World |
| 2) Mobile Transactions (Cornforth et al., 2012) | 2010 | World |
| 3) Delivery System and Method (Reeve et al., 2012) | 2011 | World |
| 4) Cellular Airtime Management (Cornforth et al., 2016) | 2012 | US, Africa |
| 5) Mobile Savings Account (Cornforth et al., 2014a) | 2012 | Africa |
| 6) Mobile Money Transfer (Cornforth et al., 2014b) | 2012 | Africa |
| 7) Optimizing financial transactions network flow (Ignatyev et al., 2015) | 2014 | Not granted |
| 8) Determining multiple users of a network enabled device (Scarr et al., 2015) | 2014 | US |
| 9) Mobile transaction validation (Cornforth et al., 2012) | 2015 | Not granted |
| 10) Digital currency conversion (Desiree, 2016) | 2015 | Not granted |

**Table 5: Patents linked to M-Pesa**
**Source: Author's analysis based on patent search.**

**Source of novelty:** Three sources underlie the novelty of claims. A first cluster formalizes mobile/mobile money practices that emerged informally in Kenya (#1, #4, #5, #6). A second solve specific problems emerging as M-Pesa has expanded as a service (#2, #8). The third describe new or future features (#3, #7, #10). The first category is the most controversial in terms of the discussion of IPR, in that a number of the patents align to common practices occurring amongst users of mobile, or mobile money in Kenya. These patents then look to formalise these practices, often through describing potential software systems which would, in the most part, support these practices. An example of this is the principal initial 2005 patent (#1) whose claims cover a software system to manage payments for MFI (micro-finance institutions), where MFI transfers might be done through mobile messaging. The presence (as outlined in the previous section) of precursor services such as G-Cash and Celpay, as well as more informal practices around mobile use within MFI suggest that much of the novelty in this patent comes out of activities that were already emerging.

**The basis for novelty claim:** In most of the patents, novelty aligns to well-known techniques or prior art, which are claimed to be novel in the presence or application to mobile money. A group savings patent (#5) which mirrors informal approaches to group savings in Kenya is an example of this. Its basis for novelty rests on the format of the system proposed - rather than saving groups distributing cash to group members, they would use systems that integrate automated mobile money to mobile airtime conversions.

**Scope of patents:** All the patents are software or process patents. The software patents describe how they would set up aspects of mobile money software and its operation. Process patents focus on techniques or processes that allow mobile network owners to make a better analysis of the mobile money network (e.g., identify fraud or improper activities within a mobile money system). In terms of scope, at least if one takes them at their broadest independent claims[4], they define extremely broad fields of novelty. Although there is little evidence of any litigation, the initial mobile account patent (#1) for example, defines a broad field many MFI system that integrates mobile money may be infringing. Similarly, the patent that outlines a technique for management of payments over airtime (#4), could suggest patent infringement in a growing number of payroll or non-profit payment systems that integrate mobile money payments.

---

[3] A patent search was made for patents that included the terms "mobile money", "mobile finance" or "M-Pesa". Patents were then filtered out to only include those relevant to the actors involved in this case. All patents found were associated with the Vodafone Group.

### 4.2.4. Software

Central to IPR is the software that enables the operation of the M-Pesa system. During the initial DFID pilot, this was contracted to a UK-based software firm, Sagentia. As mobile money interactions did not align with existing banking systems, proprietary software was written from scratch embedding several technical innovations (Wooder & Baker, 2012). Given the interactive processes of innovation during the M-Pesa pilot and early stage of the rollout (as outlined in 4.2.2.), the software was also adapted over time, dependent on the evolving business proposition and regulatory requirements (Wooder & Baker, 2012). As the service grew the responsibility for the software was directed to IBM from 2009.

During the earlier fieldwork in 2010-2011, the fact that M-Pesa software was hosted by IBM in Germany was often a subject of animated discussion by those running the service on the ground. At that time M-Pesa was having significant problems with delays and system failure, and M-Pesa agents often blamed it on the location of servers, rightly or wrongly. There was also a broader unease amongst policy experts in Kenya that large volumes of Kenyan financial transactions were being routed through Germany.

The M-Pesa G2 project (concluded in 2015) was a cross-country software project involving Vodafone, Safaricom, IBM with Huawei as the main software development partner. The aim was to upgrade the M-Pesa software. Most symbolically, the upgrade was accompanied by the first celebration of "bringing M-Pesa" home, when M-Pesa servers and software were relocated to Kenya.

It was only later that questions surrounding the G2 project came to light. Following accusations of corruption in Safaricom, auditors KPMG were commissioned to undertake a forensic investigation of recent contracts. The report was leaked to the Kenyan press in 2016 (KPMG, 2016). It reveals the cost to Safaricom for this upgrade to be $12.5m. This involved a disputed procurement process with KPMG commenting that "based on the value of other proposals…the cost agreed with Huawei is relatively high" (KPMG, 2016, p. 111). In terms of intellectual property, the KPMG report expressed concern that,

*"although the cost of this project was borne largely by Safaricom, it does not retain exclusive use of intellectual property on the project and similar mobile money projects can be rolled out in the territories without any benefits to Safaricom" (KPMG, 2016, p. 110).*

Safaricom paid the complete cost of the upgrade project. Yet the control remained within Vodafone Group.

## 5. DISCUSSION

### 5.1. M-Pesa, innovation and IPR

The ownership and control of IPR within M-Pesa has been wrought with tensions throughout its history, and this has had a significant impact on the directions of innovation. Safaricom operates the M-Pesa service in Kenya, but important decisions about directions of innovation have often been defined by the Vodafone Group in the UK. While this should not be surprising given the head-subsidiary firm relations, one can identify diverging goals and aims amongst the different stakeholders in the UK and Kenya. The Vodafone Group wishes to ensure the service maximises revenues across multiple jurisdictions. Safaricom stakeholders have looked towards cementing its market dominance and introducing further innovations. Underlying these diverging positions is the IPR, whose ownership pushes control towards the Vodafone Group. In addition, IPR ownership has led to the situation where significant transfers of funds are moving from Kenya to the UK in the form of service fees (as well as additional cost of software contracts and the 2020 purchase of the IPR).

---

[4] As would be expected for professionally written patents, they seek to define their novelty as generically as possible (to widen their scope of possible claims). A set of dependent claims within patent then provide more detailed contexts. This enables patents to cover as broad an area as possible, where dependent claims allow more refined claims innovation to protect more specific embodiments should the broader claims be found to be invalid during litigation.

How IPR came to be fully controlled by the Vodafone Group is complex, where different aspects of IPR (software, patents, branding) have different paths. The lack of attention to intellectual property during the DFID donor pilot was highly problematic in this respect. In terms of the generation of IP, innovative actions of leaders and software developers associated with Vodafone Group in the UK should not be ignored. Yet the extensive research on M-Pesa has elucidated the reality that a diversity of actors contributed to innovation, including more marginal actors in Kenya, even if such contributions are not acknowledged within the IPR.

As M-Pesa has become a global service with centrally controlled IPR it has become embedded within the strategies of the multinational parent firm, with goals to maximise its global profits. The trends of corporate expansion across multiple territories and the building of a patent portfolio to support this are in line with the literature on globalised digital firms which draw on IPR to support such intangible assets. However, one can also see the more nefarious nature of digital capitalism in the case of M-Pesa and its IPR. The extent of the various payments from Kenya to the UK is surprising given the donor-funded origins of the service. That Safaricom produced software at its own cost for Vodafone UK suggest issues of governance. The nature of patent portfolios linked to local knowledge in Kenya that are filed and controlled in London mirrors histories of British colonial power in Africa.

The recent announcement of M-Pesa IPR being returned to Africa signals hopes of a more balanced direction in the future. Yet this may simply open a new chapter in the complex relationships between different corporate subsidiaries, IPR and innovation.

### 5.2. IPR, digital innovation and inclusive innovation

As the in-depth account illustrates, many aspects in this case are context-specific. However, we can reflect more broadly on insights for IPR and digital innovations, particular for a growing range of digital products and services in the global south.

Findings reflect the broader literature on IPR regimes often leading to skewed control and detrimental impacts in the global south. As M-Pesa has shown, IPR is an important crucible for control of the directions of innovation in the digital economy, and where services expand globally it can result in new types of financial flows move from (domestic innovators in) the South to (head offices in) the North as a result of regimes of IPR.

The present IPR system is poorly set up to deal with the nature of modern innovation, and this is emphasised in the case of M-Pesa. The service embeds the involvement of a variety of actors that are important to innovative processes. Such processes and actors are, however, largely invisible in the IPR. In the case of M-Pesa, it is the managers and technologists in London and Cambridge who have the resources to go through the patent system, capture knowledge and write the code and who hence have control of the IPR. All the authors of patents found, for example, were based outside the Kenyan context but through their involvement with activities in Kenya have been able to stake claims for control of key ideas. There are also questions about how quite generic ideas are being taken out of the global 'commons' around mobile money with potential implications for who can innovate in these areas in the future.

As the literature review has outlined, there have been important critiques of IPR regimes when they look to embed public or social agendas, or when intellectual property embeds local knowledge or practices. Yet, there is scant evidence that such ideas have been so far incorporated or considered within thinking around digital innovation and this could be the focus of future more technical actions.

# 6. CONCLUSION

The case of M-Pesa in Kenya and IPR has been used to highlight important domains of control of innovation and problematic South-North transfer of resources. These problematise aspects of M-Pesa in terms of claims for development. More broadly, the different challenges of IPR discussed highlight the need for further research on inclusive innovation through the prism of IPR regimes. These are arguably not fit for purpose in the emerging era of digital capitalism.

Further research is needed to think more carefully about best practices. Academic work which opens up further cases of digital innovation in the south would be important empirically. Such research could also be more closely linked to more frameworks of global justice and economic dependency. More technical work can be done to consider how diverse actors involved in socially orientated innovations and local knowledge can be better embedded in IPR.

# REFERENCES

AFI. (2010). Enabling Mobile Money Transfer: The Central Bank of Kenya's Treatment of M-Pesa. Alliance for Financial Inclusion.

Bessen, J., & Hunt, R. M. (2007). An Empirical Look at Software Patents. Journal of Economics & Management Strategy, 16(1), 157–189.

Brain, A., Gulrajani, N., & Mitchell, J. (2014). Meeting the challenge: How can enterprise challenge funds be made to work better. UKAid.

CAK. (2020). Sector Statistics Report Q3 2019-2020. Communications Authority of Kenya.

Chang, H.-J. (2001). Intellectual Property Rights and Economic Development: Historical lessons and emerging issues. Journal of Human Development, 2(2), 287–309.

Chipchase, J., & Tulusan, I. (2006, December 1). Shared Phone Practices: Exploratory Field Research from Uganda and Beyond. CHI 2006, Boston, MA.

Cornforth, P., Downing, J., & Reeve, G. (2014a). Mobile savings account (Patent No. GB2504130A).

Cornforth, P., Downing, J., & Reeve, G. (2014b). Mobile Money Transfer (United States Patent No. US20140058928A1).

Cornforth, P., Downing, J., & Reeve, G. (2016). Cellular airtime management (United States Patent No. US9462138B2).

Cornforth, P., Reeve, G., & Downing, J. (2012). Mobile transactions (World Intellectual Property Organization Patent No. WO2012052787A1).

Davies, R., & Elgar, K. (2014). Enterprise Challenge Funds for Development: Rationales, Objectives, Approaches (Development Policy Centre Discussion Paper No. 36). Australian National University.

Desiree, C. (2016). Digital currency conversion (Patent No. GB2537683A).

Durand, C., & Milberg, W. (2020). Intellectual monopoly in global value chains. Review of International Political Economy, 27(2), 404–429.

Ebony Consulting. (2003). Financial Deepening Challenge Fund: Mid Term Review. Ebony Consulting International Ltd.

Fildes, N. (2017, March 20). Vodafone chief shows new pragmatism on partnerships. Financial Times.

Finger, J. M., & Schuler, P. (Eds.). (2004). Poor People's Knowledge: Promoting Intellectual Property in Developing Countries. Oxford University Press & World Bank.

Foster, C. G. (2013). Micro-Enterprises and Inclusive Innovation: A Study of the Kenyan Mobile Phone Sector [PhD Thesis]. University of Manchester.

Foster, C. (2014). Does Quality Matter for Innovations in Low Income Markets? The Case of the Kenyan Mobile Phone Sector. Technology in Society, 38, 119–129.
Foster, C. G., & Heeks, R. B. (2013a). Conceptualising Inclusive Innovation: Modifying Systems of Innovation Frameworks to Understand Diffusion of New Technology to Low-Income Consumers. European Journal of Development Research, 25(3), 333–355.
Foster, C. G., & Heeks, R. B. (2013b). Innovation and Scaling of ICT for the Bottom-of-the-Pyramid. Journal of Information Technology, 28(4), 296–315.
Foster, C. G., & Heeks, R. B. (2015). Policies to Support Inclusive Innovation (Development Informatics Working Paper No. 61). University of Manchester. http://www.seed.manchester.ac.uk/subjects/idpm/research/publications/wp/di/di-wp61/
Ghauri, P. N., & Rao, P. M. (2009). Intellectual property, pharmaceutical MNEs and the developing world. Journal of World Business, 44(2), 206–215.
Hall, B. H., & Harhoff, D. (2012). Recent Research on the Economics of Patents. Annual Review of Economics, 4(1), 541–565.
Haskel, J., & Westlake, S. (2017). Capitalism without Capital: The Rise of the Intangible Economy. Princeton University Press.
Heeks, R. (2020). ICT4D 3.0? Part 1—The components of an emerging "digital-for-development" paradigm. The Electronic Journal of Information Systems in Developing Countries, 86(3), 1–15.
Heeks, R. B., Amalia, M., Kintu, R., & Shah, N. (2013). Inclusive Innovation: Definition, Conceptualisation and Future Research Priorities (Development Informatics Working Paper No. 53). University of Manchester.
Heeks, R., Foster, C., & Nugroho, Y. (2014). New Models of Inclusive Innovation for Development. Innovation and Development, 4(2), 175–185.
Hughes, N., & Lonie, S. (2007). M-PESA: Mobile Money for the "Unbanked" Turning Cellphones into 24-Hour Tellers in Kenya. Innovations, 2(1–2), 63–81.
Ignatyev, O., Scarr, K., & Scharf-Katz, V. (2015). Optimizing financial transactions network flow (United States Patent No. US20150254651A1).
Irwin, D., & Porteous, D. (2005). Financial Deepening Challenge Fund Strategic Project Review. DFID & Bill and Melinda Gates Foundation.
Jiménez, A., & Zheng, Y. (2017). Tech hubs, innovation and development. Information Technology for Development, 24(1), 95–118.
KPMG. (2016). Safaricom Limited [Internal Factual Findings Report]. KPMG East Africa.
Lall, S. (2003). Indicators of the relative importance of IPRs in developing countries. Research Policy, 32(9), 1657–1680.
Mohapi, T. (2017, December 5). The curious case of Nyagaka Anyona Ouko, the mobile money transfer innovator—Part I. IAfrikan.Com. https://www.iafrikan.com/2017/12/05/is-this-kenyas-real-m-pesa-innovator-part-iii/
Morawczynski, O. (2010). Examining the Adoption, Usage and Outcomes of Mobile Money Services: The Case of M-PESA in Kenya [PhD Thesis]. University of Edinburgh.
Murdoch, T., Bowley, C., Vaughan, L., Carew, W., Hughes, N., & Lonie, S. (2007). Mobile Account Management (UKPA Patent No. WO2007020394).
Murphy, J. T., & Carmody, P. R. (2015). Africa's information revolution: Technical regimes and production networks in South Africa and Tanzania. Wiley.
Ngugi, B., Pelowski, M., & Ogembo, J. G. (2010). M-Pesa: A Case Study of the Critical Early Adopters' Role in the Rapid Adoption of Mobile Money Banking in Kenya. The Electronic Journal of Information Systems in Developing Countries, 43(3), 1–16.
Omwansa, T. (2009). M-Pesa: Progress and Prospects. Innovations, Innovations Case Discussion, 107–123.
Pagano, U. (2014). The crisis of intellectual monopoly capitalism. Cambridge Journal of Economics, 38(6), 1409–1429.

Pansera, M., & Owen, R. (2016). Innovation for de-growth: A case study of counter-hegemonic practices from Kerala, India. Journal of Cleaner Production.

Papaioannou, T. (2016). Marx and Sen on incentives and justice: Implications for innovation and development. Progress in Development Studies, 16(4), 297–313.

Peukert, A. (2017). Intellectual property and development—Narratives and their empirical validity. The Journal of World Intellectual Property, 20(1–2), 2–23.

Phillip, X. (2020, April 7). Vodacom and Safaricom in the driver's seat for M-Pesa. The Africa Report.Com. https://www.theafricareport.com/25741/vodacom-and-safaricom-in-the-drivers-seat-for-m-pesa/

Pompa, C. (2013). Understanding challenge funds. ODI, London. UK.

Porteous, D. (2006). The Enabling Environment for Mobile Banking in Africa. Bankable Frontier Associates.

Reeve, G., Maynard, J., & Cornforth, P. (2012). Delivery system and method (World Intellectual Property Organization Patent No. WO2012168735A1).

Rice, J. M. (2015). The defensive patent playbook. Berkeley Technology Law Journal, 30(4), 725–776.

Rikap, C. (In press). Amazon: A story of accumulation through intellectual rentiership and predation. Competition & Change.

Scarr, K., Ignatyev, O., & Scharf-Katz, V. (2015). Determining multiple users of a network enabled device (European Union Patent No. EP2958062A1).

Smith, A., Fressoli, M., Abrol, D., Arond, E., & Ely, A. (2016). Grassroots innovation movements. Taylor & Francis.

Srang-iam, W. (2013). Decontextualized Knowledge, Situated Politics: The New Scientific–Local Politics of Rice Genetic Resources in Thailand. Development and Change, 44(1), 1–27.

Srnicek, N. (2016). Platform Capitalism. Polity Press.

Stiglitz, J. E. (2014). Intellectual Property Rights, the Pool of Knowledge, and Innovation (Working Paper No. 20014; Working Paper Series). National Bureau of Economic Research.

Tyce, M. (2020) Beyond the Neoliberal-Statist Divide on the Drivers of Innovation: A Political Settlements Reading of Kenya's M-Pesa Success Story. World Development, 125

Unwin, P. T. H. (2009). ICT4D: Information and communication technology for development. Cambridge University Press.

Vodacom. (2019). Integrated Report for the year ended 31 March 2019. Vodacom Group Ltd.

Vodafone Group. (2017). Safaricom acquisition: Investor presentation. Vodafone Group Plc.

Vodafone Group. (2019). Annual Report 2019. Vodafone Group Plc.

WIPO. (2020). WIPO IP Statistics Data Center. World Intellectual Property Organisation.

Wooder, S., & Baker, S. (2012). Extracting Key Lessons in Service Innovation. Journal of Product Innovation Management, 29(1), 13–20.

# APPENDIX – NOTES OF CALCULATIONS OF M-PESA TRANSFERS

\* Fee moves from 10% to 5% on 1$^{st}$ Aug 2015

**Estimating M-Pesa service costs:** The details of the service agreement(s) or full data on service fees are not in the public domain. However, corporate documentation and financial reports from Vodafone Group highlight the method by which M-Pesa service costs are calculated.

Service fees a calculated based on a percentage of M-Pesa revenue (on a sliding scale as shown in the table). Service fees also include an additional component concerning indirect revenue gains of Safaricom as a result of savings in airtime commission costs. Where customers use M-Pesa to top-up directly, Safaricom saves of distribution and commission payments to local mobile vendors. In the case of Table 3, it is possible to calculate the former based upon company reports, but not the latter.

Based on the above discussion, the service fees for a financial year are calculated based on quarterly revenue splits, delayed by one quarter. This is multiplied by the service commission.

This estimation should be seen as a best-guess estimate of the M-Pesa service fee. For triangulation, several years data are available within corporate reports of the exact fee paid for services (for 2011 & 2016-2018). This suggests this estimate is a good estimate, typically an underestimate by up to €3m. Thus, the data highlights general trends around transfers rather than represent an exact figure.